\documentclass[12pt]{article}
\usepackage{amsfonts,bm}
\usepackage{amsmath}
\usepackage{cite}
\usepackage{graphicx}
\csname @addtoreset\endcsname{equation}{section}
\newcommand{\bra}[1]{\left< #1\right|}
\newcommand{\ket}[1]{\left| #1\right>}

\newcommand{\eq}{\begin{equation}}
\newcommand{\feq}{\end{equation}}
\newcommand{\eqn}{\begin{eqnarray}}
\newcommand{\feqn}{\end{eqnarray}}
\newcommand{\arr}{\begin{eqnarray*}}
\newcommand{\farr}{\end{eqnarray*}}

\newcommand{\G}{{\cal G}}

\newcommand{\cosech}{{\mathrm{cosech}}}
\font\mybb=msbm10 at 12pt
\def\bb#1{\hbox{\mybb#1}}
\def\bZ {\bb{Z}}
\def\bR {\bb{R}}

\def\bC {\bb{C}}

\begin{document}

\begin{titlepage}
\begin{flushright}
IFUM-757-FT\\
UTF-454 \\
hep-th/0407255
\end{flushright}
\vspace{.3cm}
\begin{center}
\renewcommand{\thefootnote}{\fnsymbol{footnote}}
{\Large \bf Aspects of Quantum Gravity in de~Sitter Spaces}
\vskip 25mm
{\large \bf {Dietmar Klemm$^1$\footnote{dietmar.klemm@mi.infn.it}
and Luciano Vanzo$^2$\footnote{vanzo@science.unitn.it}}}\\
\renewcommand{\thefootnote}{\arabic{footnote}}
\setcounter{footnote}{0}
\vskip 10mm
{\small
$^1$ Dipartimento di Fisica dell'Universit\`a di Milano,\\
and\\
INFN, Sezione di Milano,\\
Via Celoria 16, I-20133 Milano.\\

\vspace*{0.5cm}

$^2$ Dipartimento di Fisica dell'Universit\`a di Trento,\\
and\\
INFN, Gruppo Collegato di Trento,\\
Via Sommarive 14, I-38050 Trento.
}
\end{center}
\vspace{2cm}
\begin{center}
{\bf Abstract}
\end{center}
{\small In these lectures we give a review of recent attempts to
understand quantum gravity on de~Sitter spaces. In particular, we
discuss the holographic correspondence between de~Sitter gravity
and conformal field theories proposed by Hull and by Strominger,
and how this may be reconciled with the finite-dimensional
Hilbert space proposal by Banks and Fischler. Furthermore we review the
no-go theorems that forbid an embedding of de~Sitter spaces in
string theory, and discuss how they can be circumvented.
Finally, some curious issues concerning the thermal nature of de~Sitter
space are elucidated. 
}

\end{titlepage}

\section{Introduction and Motivation}

\label{intro}

\begin{figure}[ht]
\begin{center}
\includegraphics[width=0.7\textwidth]{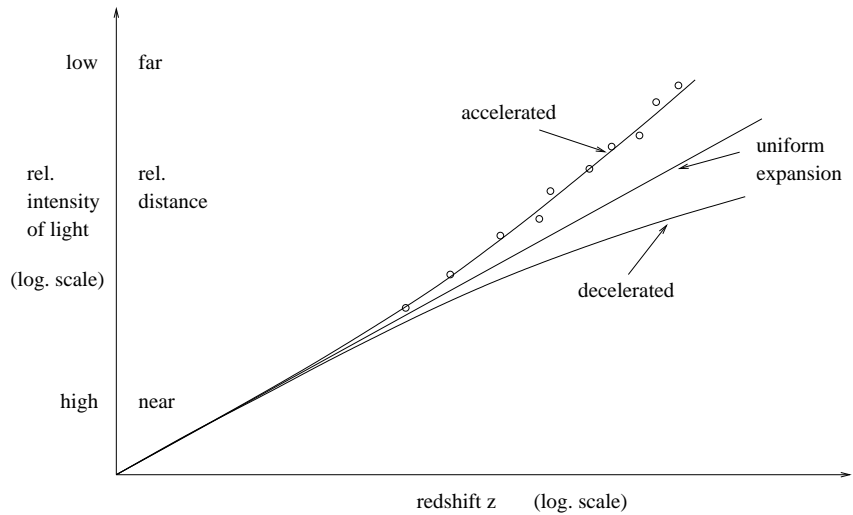}
\end{center}
\caption{\small{Relative intensity of light (corresponding to
relative distance) of type Ia supernovae measured as a function
of the redshift (corresponding to relative velocity) \cite{Riess:1998cb}.
The measurements indicate that type Ia supernovae are about 25\% dimmer
than forecast, which means that the cosmic expansion is accelerating.}
}
\label{supernovae}
\end{figure}

Recent astrophysical data coming from type Ia
supernovae \cite{Riess:1998cb} indicate that the cosmic expansion
is accelerating and point towards a small
but nonvanishing positive cosmological constant (cf.~figure
\ref{supernovae}). This means that our universe might currently
be in a de~Sitter (dS) phase.

Independent evidence for this comes from measurements of the cosmic
microwave background (CMB) anisotropy (COBE, Boomerang, WMAP),
from which we learned that the universe is spatially flat.
But in order to have a flat geometry, one needs a total energy
density much larger than the total density of ordinary matter
observed in our universe. This indicates that the major part
of the energy in the universe is a kind of ``dark energy".
The simplest and most convincing model for dark energy is a small,
positive cosmological constant $\Lambda$.
According to recent WMAP data, about 73\% of the energy density
of the universe is in a dark energy sector (with $\sim$ 22\% dark
matter and $\sim$ 4.4\% baryons) \cite{Bennett:2003bz}.

Current estimates for the energy density associated to $\Lambda$
yield
\begin{equation}
\rho_{\Lambda} = \frac{\Lambda}{8\pi G} \le \left(10^{-3}{\mathrm{eV}}\right)^4
                 \simeq 10^3\;\mathrm{eV}\cdot
                 \mathrm{cm}^{-3}\,,
\end{equation}
corresponding to a mass density
$10^{-29}\mathrm{g}/\mathrm{cm}^3$. Note that 
this value is by a factor of about $10^{123}$ too small with 
respect to the vacuum energy of the fields in the standard model,
if we take the Planck scale as a cutoff. This is the so-called
cosmological constant problem \cite{Weinberg:1988cp}. It is tempting to
envoke supersymmetry in order to solve this
problem, due to the cancellation of the vacuum fluctuations of
bosons and fermions. However, some kind of fine-tuning must
cancel only virtual-particle energies to 123 decimal places but
leave the 124-th untouched--a precision seen nowhere else in
nature. We shall not discuss the cosmological constant problem in
detail in these lectures, and refer the reader instead to the numerous
reviews \cite{Weinberg:1988cp,Carroll:2000fy}.

Apart from the current accelerating cosmic expansion,
further motivation for the interest in de~Sitter gravity comes
from the inflationary era, during which one assumes that the
universe was also described by a de~Sitter phase.

A less phenomenological and more academic problem is the realization
of the holographic principle \cite{'tHooft:gx} for spaces
more general than anti-de~Sitter (AdS) \cite{Bousso:1999cb}. For AdS gravity, we know
now that there exists a dual description in terms of
conformal field theories in one dimension lower \cite{Witten:1998qj}.
The best-understood example of this is the correspondence between
type IIB string theory on AdS$_5$ $\times$ S$^5$ and ${\cal N}=4$
super Yang-Mills theory \cite{Maldacena:1997re}. It would of course be
desirable to know further concrete realizations of the holographic
principle, for spacetimes different from AdS. As we will see
below, de~Sitter spaces represent another interesting test ground
for holography, which motivates to study dS gravity under this
point of view.

A full understanding of holography for de~Sitter gravity
seems to require an embedding of dS spaces in string theory.
However, as we will see below, there are severe obstacles to this,
resulting in no-go theorems \cite{Gibbons:1984kp,Maldacena:2000mw}.
Yet, being the most serious candidate for a complete
quantum theory of gravity, string theory should admit dS vacua,
if the astrophysical data that indicate that our universe
is currently in a de~Sitter phase, are interpreted correctly.
In this sense, dS gravity represents a big challenge for string
theory.

Last but not least, dS space possesses a cosmological event
horizon, to which one can associate a temperature and
entropy \cite{Gibbons:1977mu}. As is the case with black holes
\cite{Strominger:1996sh},
one would like to reproduce this gravitational entropy by a
microstate counting\footnote{For steps in this direction
cf.~\cite{Maldacena:1998ih,Park:1998qk}.}.

The reasons described above led to
an increasing activity on the theoretical side, with the general aim to shed
some light on quantum gravity on de~Sitter spaces \cite{Witten:2001kn},
and to embed de~Sitter space in string theory. Despite considerable effort,
these points are far from being well-understood. Different approaches do
even appear to clash with each other; for instance the proposal that
de~Sitter gravity in $D$ dimensions is dual to a Euclidean conformal
field theory in one dimension lower \cite{Strominger:2001pn} seems
to contradict the claim by Banks \cite{Banks:2000fe} and Fischler
\cite{Fischler:2000} that the Hilbert space of quantum gravity on
de~Sitter spaces is finite-dimensional. 

This review was written down while we tried to
understand these various approaches to quantum gravity on
de~Sitter spaces, and how they might be reconciled.

Our paper does not pretend to be exhaustive, rather the intention was
to elucidate some selected issues that seemed particularly interesting
to us. For a review of accelerating universes
in string theory and de~Sitter holography, that is to some extent
complementary to the material presented here, we refer the reader
to \cite{Balasubramanian:2004wx}.

Nevertheless, we still would like to mention the perturbative
results that have been obtained so far, with our apologies for any
possible omission. On the one hand these are interesting in their
own, on the other they help understanding de~Sitter more.

Past researches on de~Sitter space were directed first to uncover 
quantum field theory on a fixed dS background and semiclassical
Einstein gravity, the scheme whereby quantum fields act as source of
gravity via the expectation value of the stress tensor in a
conventional vacuum state. Certainly the problem of a choice of
vacuum, the need to understand particle detection and the stability
properties of de~Sitter space were all present in the early
investigations. We discuss these in turn. 

The richer vacuum structure of de~Sitter space was immediately
recognized. While in Minkowski
spacetime there is only one Poincar\'e invariant vacuum, for scalar
fields in de~Sitter space there is a two parameter family of
de~Sitter invariant vacua \cite{nctf}, presumably distinguished by a 
superselection parameter\footnote{It seems that no observable can
change $\alpha$.} $\alpha$. These
were studied by several authors \cite{s,Mottola:1984ar,Allen:1985ux,Burges:1984qm}
in the early days, and also more recently in \cite{Bousso:2001mw,Einhorn:2002nu};
one of these vacua is the so called Euclidean vacuum, which is uniquely selected by
a number of properties such as covariance and analyticity of the Wightman
functions, or the fact that it reduces to the ordinary Poincar\'e
vacuum as $\Lambda$ goes to zero. The other vacuum states came to be
known as the MA-states, or $\alpha$-states, after Mottola and Allen
discovered the Bogoliubov transformation relating them \cite{Mottola:1984ar,Allen:1985ux}.
They play a role especially in relation to initial perturbations in
inflationary scenarios. None of these vacua can be obtained by analytic
continuation from AdS \cite{Bousso:2001mw}.

Massless fields are more problematic; for a while it was thought that
they break de~Sitter invariance \cite{Linde:1982uu}, and in fact it was shown
that there exists no de Sitter invariant Fock representation; instead,
a two parameter family of O$(4)$-invariant states were
found \cite{Allen:1987tz}. As has been clarified by
Ford \cite{Ford:1984hs}, the physical origin of this fact is an
infrared divergence in the two-point function for such states, and
this was used by him to infer a quantum instability of de~Sitter
space. However, there exists another representation of the canonical
commutation relations leading to a de~Sitter invariant vacuum, now
known as the Kirsten-Garriga vacuum \cite{Kirsten:1993ug}.

The existence of many de~Sitter invariant vacua brings us to the
problem of particle detection and stability. One has to distinguish
between what an ``Unruh detector'' sees from the relationships among
asymptotic vacuum states at future/past
infinity \cite{Sciama:1981hr}. With one 
exception, the response of a monopole detector in the $\alpha$-states
shows an excitation rate which does not satisfy the principle of detailed
balance, but equilibrate
nevertheless \cite{Burges:1984qm,Bousso:2001mw}. The exception is the 
Euclidean vacuum, whose quantum noise is perfectly thermal and
satisfies detailed balance. 

For special values of the parameter $\alpha$ there
exist states which can be interpreted as asymptotic vacua at timelike
infinity. It is interesting that in odd-dimensional de~Sitter space
the asymptotic vacua actually coincide \cite{Bousso:2001mw}. But apart
from this case, the incoming de~Sitter vacuum for scalar particles
will decay driving a perturbative instability of the space \cite{Mottola:1984ar},
since the created particles will tend to reduce the effective
cosmological constant if the scalar mass parameter satisfies the
inequality $m^2+12\xi H^2>9H^2/4$, $H^2=\Lambda/3$ being the Hubble
constant of de~Sitter space and $\xi$ the coupling of the
scalar particle to the curvature. To our knowledge there
are no analogous results for fermions. 

Another kind of instability was discussed by
Myhrvold \cite{Myhrvold:1983hx} in an interacting $\lambda\phi^4$
theory; in this theory the spacetime curvature can make a particle to
decay into three\footnote{It is important that ``particle'' here is
  defined relative to the Euclidean vacuum.}, each one of which can
decay into three more, and so on in a runaway process. The presence of
interaction is essential for this argument to work; the back reaction
due to conformally invariant free quantum fields always gives a
semiclassically stable de~Sitter solution \cite{Myhrvold:1983hu}.  

The instabilities generated by interacting matter fields are only one
face of the problem, the other being quantum gravity itself. 
In a series of remarkable papers Tsamis and Woodard \cite{Tsamis:1992zt}
analyzed the structure of perturbative quantum gravity in de~Sitter
space in inflationary coordinates. They showed that single gravitons
can decay into two and the vacuum into three, so neither the vacuum nor
the one graviton states are stable. The instability is enhanced by
infrared divergences due to the exponential rate of expansion, which
enormously red shift all momenta to zero. They argue that, as a
result, the cosmological ``constant'' is driven to smaller values, 
thereby reducing the inflation expansion rate. 
Another source of instability come from the existence of the
Schwarzschild-de~Sitter black hole; Ginsparg and
Perry \cite{Ginsparg:1982rs} argued that this solution is a saddle
point of the Euclidean functional integral for gravity, thereby
inferring a semiclassical instability of de~Sitter space against
formation of black holes. Once formed they presumably evaporate away
on a time scale $t\simeq\Lambda^{-3/2}/G$ much shorter than that
required for another nucleation\footnote{The instability was not seen
by Abbott and Deser in their classical stability
analysis \cite{Abbott:1981ff} because they missed the Nariai
solution.}.

What can we learn from all this is hard to say, also because
{\it exact} de Sitter space is only an approximation to the real
world. We only want to note that a richer vacuum structure is an
additional degree of freedom at our disposal (for example to
incorporate inflation) and that the various instabilities may not be a
disaster in a gravitational setting, where the instability time
scale is of order of the universe lifetime. After all, we owe stars
and galaxies to gravitational clumping.\\

The lectures are organized as follows:
In order to be self-contained, we briefly review the dS geometry
in the next section. In section \ref{conscharges} we explain how
to define conserved charges associated to conformal Killing vectors
for asymptotically dS spaces. This nice idea goes back to Kastor
and Traschen \cite{Kastor:2002fu}, and circumvents the drawback that
de~Sitter space has no globally defined timelike Killing vector
to which one could associate a positive energy.
In section \ref{dSstring} we describe the no-go theorems that
prevent us from embedding dS spaces in string theory, and discuss
how these theorems can be circumvented.
The correspondence between de~Sitter gravity and Euclidean conformal
field theories is reviewed in section \ref{dSCFT}.
In \ref{finitedim}, we try to explain the motivations that led Banks
and Fischler to the proposal that the Hilbert space of quantum gravity
on dS spaces is finite-dimensional, and we indicate how this might
be reconciled with the dS/CFT correspondence.
Finally, we discuss some issues related to the thermal nature of
dS in section \ref{dSthermo}.

\section{De~Sitter Geometry}

\label{dsgeometry}

\begin{figure}[ht]
\begin{center}
\includegraphics[width=0.4\textwidth]{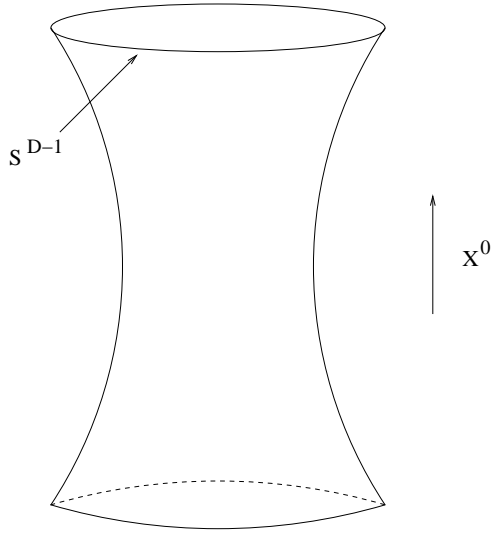}
\end{center}
\caption{\small{De~Sitter space dS$_D$ as a hypersurface in
$(D+1)$-dimensional Minkowski spacetime.}
}
\label{hyperb}
\end{figure}

Consider $(D+1)$-dimensional Minkowski space $\bR^{D+1}_1$ with coordinates
$X^A$, where $A = 0, \ldots, D$, and metric $\eta_{AB} =
{\mathrm{diag}}(-1, 1, \ldots, 1)$. dS$_D$ can then be represented
as the hypersurface
\begin{equation}
\eta_{AB} X^A X^B = \ell^2\,, \label{hyperboloid}
\end{equation}
where the constant $\ell$ is essentially the curvature radius of
de~Sitter space. The hyperboloid (\ref{hyperboloid}) is shown
in figure \ref{hyperb}.

The ``time" $X^0$ runs vertically, while
slices of constant $X^0$ are $(D-1)$-dimensional spheres S$^{D-1}$.
De~Sitter space is thus a sphere S$^{D-1}$ that contracts to reach
a minimal radius, and then reexpands. For the metric $g_{\mu\nu}$
on dS$_D$ we take the induced metric from the embedding space
$\bR^{D+1}_1$.
$D$-dimensional de~Sitter space is then a space of constant curvature,
which implies that it is also Einstein, i.~e.~, the Einstein
tensor $G_{\mu\nu}$ satisfies
\begin{equation}
G_{\mu\nu} + \Lambda g_{\mu\nu} = 0\,,
\end{equation}
with a cosmological constant given by
\begin{equation}
\Lambda = \frac{(D-2)(D-1)}{2\ell^2}\,.
\end{equation}
By construction it is clear that the isometry group of dS$_D$
is O$(D,1)$. Note that this is also the Euclidean conformal group
in $D-1$ dimensions. This will become important in section
\ref{dSCFT}.

We shall now introduce some coordinate systems that will be
useful later. If we set
\begin{equation}
X^0 = \ell\sinh\frac{\tau}{\ell}\,,
\end{equation}
we get from (\ref{hyperboloid})
\begin{equation}
(X^1)^2 + \ldots + (X^D)^2 = \ell^2\cosh^2\frac{\tau}{\ell}\,,
\end{equation}
i.~e.~, the coordinates $X^1, \ldots, X^D$ range over a $(D-1)$-sphere
with radius $\ell\cosh\frac{\tau}{\ell}$. The induced metric on (\ref{hyperboloid})
takes the form
\begin{equation}
ds^2 = -d\tau^2 + \ell^2\cosh^2\frac{\tau}{\ell} d\Omega^2_{D-1}\,,
       \label{metrglobal}
\end{equation}
where $d\Omega^2_{D-1}$ denotes the standard line element on the
unit S$^{D-1}$. The coordinates in (\ref{metrglobal}) are global,
they cover the whole manifold. From (\ref{metrglobal}) it is also
evident that dS$_D$ is a contracting and then reexpanding $(D-1)$-sphere.

\begin{figure}[ht]
\begin{center}
\includegraphics[width=0.5\textwidth]{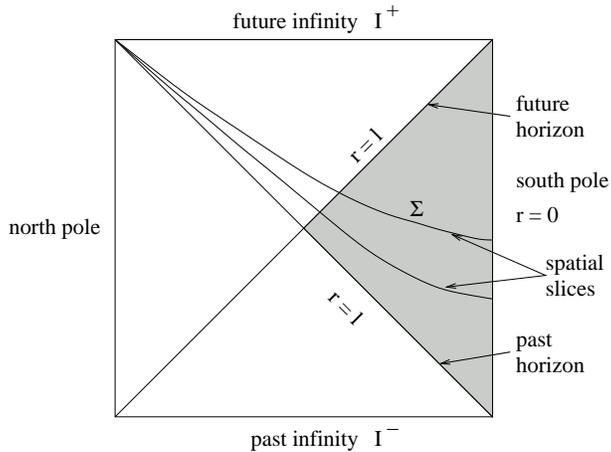}
\end{center}
\caption{\small{Carter-Penrose diagram for de~Sitter
space \cite{HawkEllis}. The future and past
horizon of a static observer sitting at the south pole $r=0$ (with
$r$ being the radial coordinate in (\ref{metrstatic})) of the
spatial $(D-1)$-sphere are shown. The static patch is shaded.
Two spatial slices $\Sigma$ of constant time in inflationary
coordinates are indicated.}
}
\label{penrose}
\end{figure}

Future inflationary coordinates $(t, x^i), i = 1, \ldots, D-1$, are
defined by 
\begin{eqnarray}
X^0 &=& \ell\sinh\frac{t}{\ell} + \frac{\vec{x}^2}{2\ell}
e^{t/\ell}\,, \nonumber \\ 
X^i &=& x^i e^{t/\ell}\,, \\
X^D &=& \ell\cosh\frac{t}{\ell} - \frac{\vec{x}^2}{2\ell} e^{t/\ell}\,. \nonumber
\end{eqnarray}
This leads to the metric
\begin{equation}
ds^2 = -dt^2 + e^{2t/\ell} d\vec{x}^2\,, \label{metrinfl}
\end{equation}
where $d\vec{x}^2$ is the $(D - 1)$-dimensional flat line element. Inflationary
coordinates cover the upper right triangle of the Carter-Penrose diagram
(cf.~figure \ref{penrose}), with $t=\infty$ and $t=-\infty$ corresponding to
future infinity and the past horizon respectively.

In order to obtain the metric in static coordinates, fix
\begin{equation}
(X^0)^2 - (X^D)^2 = r^2 - \ell^2\,. \label{hyperbola}
\end{equation}
One has then
\begin{equation}
(X^1)^2 + \ldots + (X^{D-1})^2 = r^2\,,
\end{equation}
so the coordinates $X^1, \ldots, X^{D-1}$ range over S$^{D-2}$ with
radius $r$. If we parametrize the hyperbola (\ref{hyperbola}) of fixed
$r$ by
\begin{equation}
X^0 = \sqrt{\ell^2 - r^2}\sinh\frac{t}{\ell}\,, \qquad
X^D = \sqrt{\ell^2 - r^2}\cosh\frac{t}{\ell}\,,
\end{equation}
then the induced metric on the hypersurface (\ref{hyperboloid})
is given by
\begin{equation}
ds^2 = -\left(1-\frac{r^2}{\ell^2}\right)dt^2 + \left(1-\frac{r^2}{\ell^2}
       \right)^{-1}dr^2 + r^2 d\Omega_{D-2}^2\,, \label{metrstatic}
\end{equation}
where $d\Omega^2_{D-2}$ is the round metric on the unit S$^{D-2}$.
The line element (\ref{metrstatic}) becomes singular for $r = \ell$,
which is a cosmological event horizon to which one can associate
a temperature and entropy \cite{Gibbons:1977mu}
\begin{equation}
T = \frac{1}{2\pi\ell}\,, \qquad S = \frac{A_{\mathrm{Hor}}}{4G}\,.
\end{equation}
Further useful parametrizations can be found in
\cite{Klemm:2001ea,Spradlin:2001pw}. 

The Carter-Penrose diagram of de~Sitter space is shown in figure
\ref{penrose}.

We noted above that de~Sitter space has topology $\bR\times$ S$^{D-1}$
and the noncompact isometry group O$(D,1)$. If one
requires dS physics to be maximally symmetric then this
and elliptic de~Sitter space\footnote{This is obtained from dS
  through the antipodal identification $X^A\to-X^A$ of the embedding
  coordinates, and is a maximally symmetric, non time-orientable
  manifold with isometry group SO$(D,1)$\cite{Parikh:2002py}.} are the only
possibilities. As was observed by Witten \cite{Witten:2001kn}, maximal
symmetry with non-compact groups is not welcome in a theory whose
content is restricted to a finite-dimensional Hilbert space, and it was
also strongly suggested \cite{Goheer:2002vf} that the $O(D,1)$ symmetry
must be spontaneously broken in the vacuum. The case for less
symmetric versions of de~Sitter (and anti-de~Sitter) space was then
proposed and defended in an elegant manner by
McInnes \cite{McInnes:2003xm,McInnes:2003ya}.\\
Having de~Sitter spaces with less symmetry is therefore a good avenue
for quantum gravity, and a way to achieve this is taking quotients by
discrete groups of isometries. In fact, if a space $M$ with symmetry
group $\G$ is quotiented by a discrete subgroup $\Gamma$, acting
without fixed points, then the isometries of $M$ that are also
isometries of $M/\Gamma$ are those in the normalizer $N(\Gamma)$
of $\Gamma$, that is those $g\in\G$ such that $g\gamma
g^{-1}\in\Gamma$ for any $\gamma\in\Gamma$. But since $\Gamma$ acts
trivially on $M/\Gamma$, the true isometry group is actually
$N(\Gamma)/\Gamma$, which is even smaller than $N(\Gamma)$, and
in general far smaller than the original
group. In this way the symmetry is broken.\\
How much symmetry must be broken is unknown, but it seems fair enough
to assume that a generic FLRW-like symmetry should remain, namely the
universe should be spatially isotropic and homogeneous. This singles out
the group $\Gamma=\bZ_2$ generated by the antipodal map
$(X^0,X^1,\dots,X^D)\to(X^0,-X^1,\dots,-X^D)$, which
gives a de~Sitter universe with spatial topology $\bR {\mathrm{P}}^{D-1}$. The
metric remains the same, given by \eqref{metrglobal}. For $D-1=2k$
this space in not orientable, and $\bR {\mathrm{P}}^{1+4k}$ is not a spin
manifold, so dS$(\bR {\mathrm{P}}^3)$ and dS$(\bR {\mathrm{P}}^7)$ seem
to be the only two available alternatives to a spherically based
de~Sitter space dS$({\mathrm{S}}^n)$,
at least for sufficiently low dimensions. The universe with 
$\bR{\mathrm{P}}^3$ 
spatial section was actually the preferred choice for de~Sitter
himself, for reasons that he left unexplained.\\
From the general rule given above, it is not difficult to see that the
isometry group of projective de~Sitter space is the compact group
$\bZ_2\times {\mathrm{O}}(4)/\bZ_2$, where the first $\bZ_2$ factor is
generated 
by the matrix $\mathrm{diag}(-1,1,1,1,1)$ and the second $\bZ_2$
by the matrix $\mathrm{diag}(1,-1,-1,-1,-1)$. Apart from a discrete
symmetry exchanging the future with the past, we see that the space is
spatially homogeneous and isotropic with a compact symmetry group,
and admits now finite-dimensional unitary representations. In view of
these facts, it seems not easy to decide which is the ``correct
version'' of de Sitter space (see \cite{McInnes:2003xm} for a detailed
discussion).

\section{Conserved Charges}

\label{conscharges}

In order to make sense of a (still to be discovered) theory of quantum gravity
on de~Sitter spaces, or even of dS gravity classically, one
would like to have a definition of energy in de~Sitter spacetime, and
this energy 
should be positive definite. Usually, a definition of energy is provided
by the Abbott and Deser construction \cite{Abbott:1981ff}, where one considers
spacetimes that asymptote at infinity to a fixed background, which
has a certain number of Killing vectors. For each background Killing vector
there exists then a conserved current, and the corresponding conserved charge
can be expressed as a boundary integral at spatial infinity.
Energy is then the charge associated to the time translation Killing vector.
If the considered background is Minkowski or AdS space, one can use
supersymmetry to show that this energy is positive
definite \cite{Witten:mf,Breitenlohner:bm}.
This is however not possible for dS space, as the latter is not a
supersymmetric vacuum state of ordinary supergravity
theories\footnote{Note that 
de~Sitter superalgebras exist \cite{Pilch:aw,McKeon:2003gm}, but they do
not have unitary 
highest weight representations. Accordingly, the corresponding
supergravity theories 
(that admit supersymmetric dS vacua) have ghosts \cite{Pilch:aw}.}.
Another, related reason, why one does not expect a positive energy theorem to
hold for dS gravity is the absence of a globally defined timelike
Killing vector field. 
For instance the Killing vector $\partial_t$ of the metric
(\ref{metrstatic}) is 
timelike inside the horizon $r = \ell$, but becomes spacelike for $r > \ell$.
Correspondingly, the conserved charge associated to $\partial_t$
receives negative 
contributions from matter or gravitational fluctuations outside the
horizon \cite{Abbott:1981ff}. This is related to the fact that
actually $\partial_t$ generates a boost in the isometry group, and a boost
always has fixed points. More generally, all conjugacy
classes\footnote{A conjugacy class in a group is a set of elements of
  the form $g\gamma g^{-1}$. Different conjugacy classes are
  necessarily disjoint as they define an equivalence relation in the
  group.} in the
de~Sitter groups O$(4k,1)$ are ambivalents, which means that every
group element is in the same class as its inverse. It then follows
that every infinitesimal generator $H$ can be transformed into its
negative, that is, a group element $G$ can be found such that
$GHG^{-1}=-H$. In any unitary representation no positive definite
operators can then be found, and it turns out that if $H$ represents
the energy operator, the element $G$ reversing the sign is the
rotation sending a point to its antipode.\\
A further difficulty consists in the fact that in global coordinates
(\ref{metrglobal}) 
the spatial slices of dS$_D$ are closed $(D-1)$-spheres, so they have
no spatial
infinity at which to define an ADM-like expression for the mass.
In spite of these objections, Kastor and Traschen
\cite{Kastor:2002fu} were 
able to prove a positive energy theorem for asymptotically de~Sitter spaces.
The difficulties stated above are thereby circumvented as follows:
First of all, 
energy is defined as the conserved charge associated to a globally
timelike {\it conformal} 
Killing vector rather than to the time translation Killing vector
$\partial_t$ of (\ref{metrstatic}) (which, as
we said, is not globally timelike). Similar to Witten's proof of
positive energy 
for asymptotically flat spaces \cite{Witten:mf}, one can then use a spinor
construction to show that this charge is positive\footnote{Shiromizu et
al.~\cite{Shiromizu:2001bg} arrive independently at the association of a
positive conserved charge with a globally timelike conformal Killing
vector, but their 
line of reasoning is different from that of Kastor and Traschen: Positivity
of the charge is derived starting from the ordinary positive energy theorem
in a conformally related asymptotically flat spacetime.}.
Thereby the supercovariant derivative
\begin{equation}
\hat{\nabla}_{\mu} = \nabla_{\mu} + \frac{i}{2\ell}\gamma_{\mu}\,,
\end{equation}
that appears in the dS supergravities of \cite{Pilch:aw}, plays an essential role.
This means that dS supergravity, though facing non-unitary problems
when quantized (at 
least perturbatively), is nevertheless useful for deriving classical results.
Of course one must specify an asymptotic region where the mass is to be defined.
As explained above, this is problematic in de~Sitter space, where the
spatial slices 
are compact (if we use global coordinates), and thus there is no spatial
infinity. To overcome this, the authors of \cite{Kastor:2002fu} considered metrics
that approach asymptotically (in the far future) that of dS spacetime
in inflationary 
coordinates (\ref{metrinfl}). The spatial slices, two of which are shown in the
Carter-Penrose diagram \ref{penrose}, are then planes, and the point at the upper
left hand corner of the diagram can be regarded as spatial infinity.

Let us describe slightly more in detail the construction of \cite{Kastor:2002fu},
which is for $D=4$, so we specialize to this case in the rest of this section.
The generalization to arbitrary dimension is straightforward.
The isometry group of dS$_4$ is SO(4,1), and the conformal group SO(4,2).
The five additional generators, i.~e.~, the conformal Killing vectors, are simply
the projections of the translational Killing vectors $\xi^{(A)} =
\partial/\partial X^A$ 
of the embedding space onto the hyperboloid (\ref{hyperboloid}). The particular
linear combination $\zeta = \xi^{(0)} - \xi^{(4)}$ reads
\begin{equation}
\zeta = e^{t/\ell}\partial_t\,,
\end{equation}
and is globally timelike and future directed. It is the conserved
charge $Q_{\zeta}$ 
associated with $\zeta$ that is non-negative. For a general spacetime
that asymptotes 
to (\ref{metrinfl}) in the far future, $Q_{\zeta}$ can be calculated in the
following way: Denote the metric of the general, asymptotically dS spacetime by
$g_{\mu\nu}$, and that of the background (\ref{metrinfl}) by $\tilde{g}_{\mu\nu}$.
Define the deviation
\begin{equation}
h_{\mu\nu} = g_{\mu\nu} - \tilde{g}_{\mu\nu}\,,
\end{equation}
which is small near spatial infinity. Define furthermore
\begin{eqnarray}
H^{\mu\nu} &=& h^{\mu\nu} - \frac 12 \tilde{g}^{\mu\nu} h\,, \nonumber \\
K^{\alpha\beta\gamma\delta} &=& \frac 12 [\tilde{g}^{\alpha\delta}H^{\beta\gamma} +
                                   \tilde{g}^{\beta\gamma}H^{\alpha\delta} -
                                   \tilde{g}^{\alpha\gamma}H^{\beta\delta} -
                                   \tilde{g}^{\beta\delta}H^{\alpha\gamma}]\,, \nonumber \\  
{\cal B}^{\alpha\beta} &=& (\tilde{\nabla}_{\gamma}K^{\alpha\beta\delta\gamma})\zeta_{\delta}
          -K^{\alpha\beta\delta\gamma}\tilde{\nabla}_{[\gamma}\zeta_{\delta]}\,,  
\end{eqnarray}
where $\tilde{\nabla}$ denotes the connection of the de~Sitter
background. If one chooses 
a spatial surface $\Sigma$ of the kind discussed above, with
boundary $\partial\Sigma$, the conserved charge associated to $\zeta$ reads
\begin{equation}
Q_{\zeta} = \frac{1}{8\pi G}\int_{\partial\Sigma} {\cal B}^{\mu\nu} dS_{\mu\nu}\,,
            \label{Qzeta}
\end{equation}
where $dS_{\mu\nu}$ denotes the background area element.
As an example, consider the Reissner-Nordstr\"om-de~Sitter solution
\begin{equation}
ds^2 = -F(R)dT^2 + \frac{dR^2}{F(R)} + R^2 d\Omega_2^2\,, \qquad
F(R) = 1 - \frac{2m}{R} + \frac{q^2}{R^2} - \frac{R^2}{\ell^2}\,. \label{RNdS}
\end{equation}
In inflationary-type coordinates, (\ref{RNdS}) reads\footnote{The coordinate
transformation connecting (\ref{RNdS}) and (\ref{RNdS_infl}) is given by
$R=\sqrt{U(r')}r'$, $T=t-\int W(r')dr'$, $W(r')=\frac{r'}{\ell}U(r')/(\frac{{r'}^2}{\ell^2}
U(r')-V(r'))$, $r'=a(t)r$.}
\begin{equation}
ds^2 = -V(ar)dt^2 + a^2 U(ar)(dr^2 + r^2 d\Omega_2^2)\,, \label{RNdS_infl}
\end{equation}
where $a = \exp(t/\ell)$ and
\begin{equation}
V(ar) = \frac{\left[1 - \frac{m^2 - q^2}{4a^2r^2}\right]^2}{\left[\left(1 +
        \frac{m}{2ar}\right)^2 - \frac{q^2}{4a^2r^2}\right]^2}\,, \qquad
U(ar) = \left[\left(1 + \frac{m}{2ar}\right)^2 - \frac{q^2}{4a^2r^2}\right]^2\,. 
\end{equation}
The Carter-Penrose diagrams as well as a detailed study of this
metric in relation to the dS/CFT duality can be found in the
interesting paper \cite{Astefanesei:2003gw}. 
Using (\ref{Qzeta}), one obtains for the charge associated to the
background conformal Killing vector $\zeta = a(t)\partial_t$,
\begin{equation}
Q_{\zeta} = a(t)\frac mG\,.
\end{equation}

One might wonder about the time-dependence of $Q_{\zeta}$, since, as was explained
above, it is a conserved charge. The reason is that there is a nonzero flux of the
spatial components of the conserved current ${\cal C}^{\alpha} =
\tilde{\nabla}_{\beta}{\cal B}^{\alpha\beta}$ through spatial infinity.

We close this section with the remark that the existence of a conserved charge
associated to a conformal Killing vector, and the fact that this charge can
be shown to be positive semidefinite using the covariant derivative of
dS supergravity seems to suggest that there are some residues of conformal
supersymmetry in de~Sitter.

\section{De~Sitter Space in String Theory}

\label{dSstring}

There are several reasons that make it desirable to embed de~Sitter spaces
in string theory. First, a full understanding of holography for dS gravity
seems to require such an embedding. Second, being the most serious candidate
for a complete quantum theory of gravity, string theory should admit dS vacua,
if the astrophysical data that indicate that our universe
is currently in a de~Sitter phase, are interpreted correctly.
However, dS vacua seem to be forbidden in string theory, due to some
no-go theorems \cite{Gibbons:1984kp,Maldacena:2000mw}. In this section we will
review them and discuss some possible ways in which these theorems can be
circumvented.

\subsection{No-Go Theorems}

The first no-go theorem is due to Gibbons \cite{Gibbons:1984kp}. (For a
more recent review cf.~\cite{Gibbons:2003gb}). He considered
warped compactifications $M = X \times_w Y$, where $M$ and $X$ denote
$N$- and $d$-dimensional Lorentzian manifolds respectively, and $Y$
is a compact $(N-d)$-dimensional Euclidean space. In local coordinates
$x^M$ which split as $x^{\mu}$ for $X$ and $y^m$ for $Y$, the metric on $M$
reads
\begin{equation}
g_{MN}\, dx^M dx^N = w^2(y)\, g_{\mu\nu}\, dx^{\mu} dx^{\nu} + g_{mn}\, dy^m dy^n\,,
\label{warpedcomp}
\end{equation}
where $w(y)$ is the warp factor. The form of the line element (\ref{warpedcomp}) is
restricted by the fact that for all pure supergravity models the bosonic
energy-momentum tensor $T_{MN}$ satisfies the strong energy condition, i.~e.~,
\begin{equation}
\left(T_{MN} - \frac 1{N-2}g_{MN}{T^L}_L\right)V^M V^N \ge 0
\end{equation}
for all non-spacelike vectors $V^M$. If we use the Einstein equations in $M$,
\begin{equation}
R_{MN} = 8\pi G_N \left(T_{MN} - \frac 1{N-2}g_{MN}{T^L}_L\right)\,,
\end{equation}
we obtain
\begin{equation}
R_{MN} V^M V^N \ge 0\,,
\end{equation}
or
\begin{equation}
R_{00} \ge 0 \label{R00}
\end{equation}
in local coordinates. Physically this means that locally gravity is
attractive. Applied to cosmology, it implies that the acceleration
of the universe is always negative. This can be seen as follows:
If $Y$ is compact without boundary and $w(y)$ is smooth and nowhere
vanishing, (\ref{R00}) implies ${}^X\! R_{00} \ge 0$, where ${}^X\! R_{\mu\nu}$
denotes the Ricci tensor of $X$ \cite{Gibbons:2003gb}.
Now take e.~g.~$X$ to be Einstein, ${}^X\! R_{\mu\nu} = \lambda g_{\mu\nu}$.
One has then $\lambda \le 0$, so that de~Sitter space is excluded.

The second no-go theorem goes back to Maldacena and
Nu\~{n}ez \cite{Maldacena:2000mw}. They also consider warped compactifications
from $N$ to $d$ dimensions. The assumptions that go into the theorem are
\begin{itemize}
\item Higher curvature corrections (like those coming from integrating out massive
string modes) are absent in the gravitational action
\item The scalar potential is nonpositive
\item Massless fields come with positive kinetic terms (note that this assumption
is violated in Hull's II* theories \cite{Hull:1998vg})
\item The $d$-dimensional effective Newton constant is finite
\item The only possible singularities are such that the warp factor $w(y)$
goes to zero at the singularity
\end{itemize}
These assumptions imply that there is no compactification to de~Sitter
space \cite{Maldacena:2000mw}.

\subsection{How to circumvent them}

Of course a no-go theorem is no better than the assumptions that go into it. There
are several ways to circumvent the above theorems, e.~g.~by
\begin{itemize}
\item Considering non-compact internal manifolds \cite{Gibbons:2001wy},
like e.~g.~hyperbolic spaces. ``Compactification" of ten- or eleven-dimensional
supergravity on such spaces gives rise to lower-dimensional supergravities
with non-compact gaugings, which indeed admit dS vacua. The problem here is
that the $d$-dimensional Newton constant $G_d$, which is related to the
Newton constant $G_N$ in $N$ dimensions by $G_d = G_N/V_Y$, is zero, because
the volume $V_Y$ of the internal manifold is infinite. This indicates that gravity
is not localized in the large extra dimensions of the internal space.
\item Coupling supergravity to super-matter \cite{Fre:2002pd}. Generically,
in such matter-coupled theories the strong-energy condition is violated,
and the scalar potential can be positive, so that both Gibbon's theorem and
that by Maldacena/Nu\~{n}ez can be circumvented, and dS solutions are possible.
It is however not clear how to embed these solutions in ten-or eleven-dimensional
supergravity.
\item Considering Hull's II* theories \cite{Hull:1998vg}. The IIB* theory, for
instance, admits a dS$_5$ $\times$ H$^5$ vacuum. Unfortunately, it is not clear
if these theories are well-defined, because the kinetic terms
of all RR fields have the wrong sign, which might lead to ghosts.
\item Including localized sources with negative tension \cite{Giddings:2001yu}.
Note that such sources are present in string theory (e.~g.~orientifold planes).
The no-go theorem of \cite{Maldacena:2000mw} states that in the absence of
localized sources there can be no NS or RR fluxes, which are necessary for
a nonconstant warp factor. (A constant warp factor does not allow
dS compactifications \cite{Maldacena:2000mw}).
\item Including $\alpha'$ or quantum corrections to the leading order
supergravity Lagrangian \cite{Kachru:2003aw}. The authors of \cite{Kachru:2003aw}
consider IIB compactifications with nontrivial NS and RR three-form fluxes,
and thus, as was explained above, they also need localized sources.
Furthermore, they include nonperturbative quantum corrections to the superpotential
for the Calabi-Yau moduli, which is generated in the presence of nonzero fluxes.
\item Including the tachyon. This violates the strong energy condition. The rolling
of the tachyon down its potential, away from the false (perturbative) vacuum, has
been proposed as a mechanism for inflation. An accelerating universe is possible
in such a scenario, but there are several shortcomings as a mechanism for inflation.
A review of tachyon cosmology can be found in \cite{Gibbons:2003gb}.
\end{itemize}

\subsection{Spacelike Branes}

There is yet another way to overcome the above no-go theorems, namely
by allowing the size of the internal manifold to change in time.
(Note that the metric (\ref{warpedcomp}) is static).
This idea has been used by Townsend and Wohlfarth \cite{Townsend:2003fx} to
obtain FLRW universes in string theory that show a transient
phase of acceleration. This makes these solutions phenomenologically
interesting, although they are not really dS spaces\footnote{After
  the Townsend and Wohlfarth paper, other interesting accelerating
  cosmologies in this vein were found
  in \cite{Ohta:2003pu,Roy:2003nd,Chen:2003dc,Neupane:2003cs}.}.
Actually the accelerating cosmologies presented in \cite{Townsend:2003fx}
are a special case of so-called spacelike branes
(S-branes) \cite{Gutperle:2002ai}. Before we come to a closer description
of this type of branes, let us present the general idea
following \cite{Emparan:2003gg}. We start from the higher-dimensional
action
\begin{equation}
I = \frac 1{16\pi G_{4+n}}\int d^{4+n}x \sqrt{-g}\left(R - \frac 1{2\cdot 4!}F^2_{[4]}
    \right)\,,
\end{equation}
and let the geometry be a warped product of a four-dimensional spacetime and
an internal compact Einstein manifold $Y_{\sigma,n}$ with
${}^Y\! R_{mn} = \sigma (n-1) g_{mn}$, $\sigma = 0, \pm 1$,
\begin{equation}
ds^2 = e^{-n\psi(x)} g_{\mu\nu}(x)\, dx^{\mu} dx^{\nu} + e^{2\psi(x)}
       g_{mn}\, dy^m dy^n\,. \label{KKtimedep}
\end{equation}
The field strength is taken as $\ast F_{[4]} = b\,{\mbox{vol}}(Y_{\sigma,n})$.
This leads upon reduction on $Y_{\sigma,n}$ to the four-dimensional action
\begin{equation}
I = \frac 1{16\pi G_4}\int d^4 x \sqrt{-g}\left(R - \frac{n(n+2)}2 (\nabla\psi)^2
    - V(\psi)\right)\,, \label{4dact}
\end{equation}
with the potential
\begin{equation}
V(\psi) = \frac{b^2}2 e^{-3n\psi} - \sigma n(n-1)e^{-(n+2)\psi}\,.
\end{equation}
A particularly interesting case appears for $n=7$, which corresponds to a
compactification of eleven-dimensional supergravity down to four dimensions.
For vacuum solutions without flux ($b=0$), one needs $\sigma < 0$ in order
to have a positive potential (which is necessary to get an accelerated cosmic
expansion in four dimensions). Thus, for $b=0$, the internal manifold
$Y_{\sigma,n}$ must have negative scalar curvature. In particular, it could
be a space of constant negative curvature, obtained by identifying hyperbolic
$n$-space under the action of a freely acting discrete subgroup of its
SO$(1,n)$ isometry group. This is the case considered in \cite{Townsend:2003fx}.
For flux compactifications, $b \neq 0$, there is always a positive contribution
to the potential, and also $\sigma \ge 0$ is possible.

The equations of motion following from (\ref{4dact}) admit the FLRW
solutions \cite{Emparan:2003gg}
\begin{equation}
g_{\mu\nu}\, dx^{\mu} dx^{\nu} = - a^6(t)\, dt^2 + a^2(t)\, d{\vec x}^2_3\,,
                                 \label{FLRW}
\end{equation}
where $d{\vec x}^2_3$ denotes the three-dimensional flat metric, and the scale
factor is given by
\begin{equation}
a(t) = e^{B + n\psi/2}\,,
\end{equation}
with
\begin{eqnarray}
B(t) &=& -\frac 13\log\left(b\sqrt{\frac{n-1}{6(n+2)}}\cosh 3(t-t_0)\right)\,,
         \nonumber \\
\psi(t) &=& \frac 1{n-1}\log\gamma(t) - \frac 3{n-1}B(t)\,,
\end{eqnarray}
\begin{displaymath}
\gamma(t) = \left\{ \begin{array}{l@{\qquad}l} \beta\,\cosech[(n-1)\beta |t|]\,, &
            \sigma = -1\,, \vphantom{\displaystyle\frac.1} \\
            \exp[(n-1)\beta t]\,, & \sigma = 0\,, \vphantom{\displaystyle\frac.1} \\
            \beta\,\mathrm{sech}[(n-1)\beta t]\,, & \sigma = +1\,, \end{array}\right.
\end{displaymath}
and
\begin{displaymath}
\beta = \frac 1{n-1}\sqrt{\frac{3(n+2)} n}\,.
\end{displaymath}
Here, $t_0$ denotes an integration constant.
The proper time $\tau$ of a four-dimensional observer is defined by
$d\tau = a^3(t)\, dt$. A closer analysis of the above solution shows
that the condition for acceleration, $d^2 a/d\tau^2 > 0$, is satisfied
near the turning point of the radion $\psi$, which starts at $\psi \to +\infty$
with large kinetic energy and runs up the potential \cite{Emparan:2003gg}.
Note that, in this model, dark energy would be represented by the radion potential.
A detailed study of (\ref{4dact}) (with $n=7$ and added dark matter)
from a phenomenological point of view was presented in \cite{Gutperle:2003kc}. There, it
was found that the model might not be phenomenologically viable, because the
Compton wavelengths $m_{\mathrm{KK}}^{-1}$ of the Kaluza-Klein particles
are of the same order as the size of the observable part of the universe,
so that our universe would be effectively eleven-dimensional.

If we lift (\ref{FLRW}) to $4+n$ dimensions using (\ref{KKtimedep}), we obtain
\begin{equation}
ds^2 = -e^{6B + 2n\psi}dt^2 + e^{2B}d{\vec x}^2_3 + e^{2\psi} g_{mn}\, dy^m dy^n\,.
       \label{SM2}
\end{equation}
This solution, which was found in \cite{Chen:2002yq}, is an example of a spacelike
brane (if $g_{mn}$ is the metric on a space of constant negative curvature).
In particular, for $n=7$, (\ref{SM2}) represents an S2-brane of M-theory
(SM2-brane). An S-brane is a topological defect for which all of its longitudinal
dimensions are spacelike, and therefore exists only for a moment of time.
In the example above, the metric on the brane is given by $d{\vec x}^2_3$.
The symmetry group of an Sp-brane in $N$ dimensions is
ISO$(p+1)$ $\times$ SO$(1,N-p-2)$, where ISO$(p+1)$ is the group of motions on the
Euclidean world volume of the brane, and SO$(1,N-p-2)$ represents the isometry
group of the hyperbolic $(N-p-2)$-manifold, into which the space transverse to
the brane is sliced. Similar to Dp-branes, which are hypersurfaces where open
strings can end, there exist SDp-branes, in which the time coordinate obeys
a Dirichlet boundary condition.

For further literature on spacelike branes, we refer the reader
to \cite{Kruczenski:2002ap} and references therein.

\section{The dS/CFT Correspondence}

\label{dSCFT}

First evidence for a correspondence between de~Sitter gravity in $D$
dimensions and conformal field theories in $D-1$ dimensions was
given by Hull \cite{Hull:1998vg}\footnote{For related work
cf.~\cite{Park:1998qk,Park:1998yw}.}, who considered so-called
IIA* and IIB* string theories, which are obtained by T-duality on a
timelike circle from the IIB and IIA theories respectively. The type IIB*
theory admits E4-branes, which are the images of D4-branes under
T-duality along the time coordinate of the brane. The E4-branes interpolate
between Minkowski space at infinity and dS$_5$ $\times$ H$^5$ near the
horizon, where H$^5$ denotes hyperbolic space. The effective action
describing E4-brane excitations is
a Euclidean $D=4$, ${\cal N}=4$ U$(N)$ super Yang-Mills theory,
which is obtained from SYM in ten dimensions by reduction on a six-torus
with one timelike circle. This leads to a duality between type IIB*
string theory on dS$_5$ $\times$ H$^5$
and the mentioned Euclidean SYM theory \cite{Hull:1998vg}.
Unfortunately this example is pathological, because both theories have
ghosts\footnote{Note however that Euclidean super Yang-Mills theory
can be twisted to obtain a well-defined topological field theory in
which the physical states are the BRST cohomology classes \cite{deMedeiros:2001kx}.
According to \cite{Hull:1998vg}, this should correspond to a twisting
of the type IIB* string theory, with a topological gravity
limit.}.

Later Strominger proposed a more general holographic duality
relating quantum gravity on dS$_D$ to a conformal field theory residing
on one of the conformal boundaries of dS$_D$ \cite{Strominger:2001pn}.
One of the reasons that lead to this proposal is the fact that the
isometry group O$(D,1)$ of dS$_D$ coincides with the Euclidean conformal
group in $D-1$ dimensions. If physical states of quantum gravity form
a nontrivial representation of this group, this suggests that quantum
gravity on dS$_D$ is equivalent to a conformal field theory in one
dimension lower. Strominger argued that in general
this CFT may be non-unitary, with operators having complex conformal weights,
if the dual bulk fields are sufficiently massive. We mention that an
important check of the suggested dS/CFT correspondence came from the
calculation of conformal anomalies \cite{Nojiri:2001mf}, from
applications to the problem of quantum creation of de Sitter universes
\cite{Nojiri:2001qq} and from black hole physics \cite{Cvetic:2001bk}.

Further support for a dS/CFT correspondence comes from
three-dimensional de~Sitter gravity, where more quantitative
predictions can be made. In the absence of matter, dS$_3$ gravity
can be written as an SL$(2,\bC)$ Chern-Simons
theory \cite{Witten:1988hc}, with action
\begin{equation}
I = \frac{is}{4\pi}\int{\mathrm{Tr}}(A\wedge dA + \frac 23
    A\wedge A\wedge A) -
    \frac{is}{4\pi}\int{\mathrm{Tr}}(\bar{A}\wedge d\bar{A} + \frac 23
    \bar{A}\wedge \bar{A}\wedge \bar{A})\,, \label{CSaction}
\end{equation}
where
\begin{equation}
s = -\frac{\ell}{4G}\,,
\end{equation}
and
\begin{equation}
A = A^a\tau_a = \left(\omega^a + \frac i\ell e^a\right)\tau_a\,, \qquad
\bar A = {\bar A}^a\tau_a = \left(\omega^a - \frac i\ell e^a\right)\tau_a\,,
\quad a = 0,1,2\,, \label{CSconn}
\end{equation}
$e^a$ denoting the dreibein,
$\omega^a = \frac 12 \epsilon^{abc}\omega_{bc}$
the spin connection, and the $\tau_a$ are SL$(2,\bR)$ generators.

Chern-Simons theory is known to reduce to a WZNW model in presence of
a boundary \cite{Elitzur:1989nr}\footnote{As we saw in section \ref{dsgeometry},
de~Sitter space has two conformal boundaries rather than one.
In the reduction from
the Chern-Simons theory (\ref{CSaction}) to a WZNW model carried out
in \cite{Cacciatori:2001un}, only the past boundary is considered. This
might be motivated for instance if one is interested in spacetimes that are
asymptotically dS in the past but not necessarily in the future. This occurs
e.~g.~for configurations that finish in a big crunch, in which case there exists no
future boundary. Below we will comment on this more in detail.}.
The boundary conditions for
asymptotically past de~Sitter spaces \cite{Strominger:2001pn} provide then
the constraints for a Hamiltonian reduction from the WZNW model
to Liouville field theory, leading to the
action \cite{Cacciatori:2001un,Klemm:2002ir}
\begin{equation}
I = \frac{1}{4\pi}\int \sqrt g d^2x \left[\frac 12 g^{ij}\partial_i
    \Phi\partial_j\Phi + \frac{\lambda}{2\gamma^2}\exp (\gamma\Phi)
+\frac{Q}{2}\,R\Phi\right]\,,
\label{Liouvilleact}
\end{equation}
where the "cosmological constant" $\lambda$, the coupling constant
$\gamma$ and the background charge $Q$ are given by
$\lambda = 16{\ell}^{-2}$, $\gamma = \sqrt{8G/\ell}$ and
$Q = \sqrt{\ell/2G}$ respectively. The Liouville model (\ref{Liouvilleact})
is defined on the past conformal boundary ${\cal I}^-$ of dS$_3$.
As is well-known, Liouville theory has a classical central charge $c=12/\gamma^2$,
which reproduces correctly the central charge $c=3\ell/2G$ that appears
in the asymptotic symmetry algebra of dS$_3$
gravity \cite{Strominger:2001pn,Klemm:2001ea}.

One can compute the Liouville field corresponding to the
Schwarz\-schild-de~Sitter solution\footnote{We focus for the moment on the
case $\mu>0$. The case $\mu<0$ will be considered later.}
\begin{equation}
ds^2 = -\left(\mu - \frac{r^2}{\ell^2}\right) dt^2 + \left(\mu - \frac{r^2}{\ell^2}
       \right)^{-1} dr^2 + r^2 d\phi^2\,, \label{SSdS}
\end{equation}
with the result \cite{Klemm:2002ir}
\begin{equation}
e^{\gamma\Phi} = \mu\left[e^{\frac i2 \sqrt{\mu} (z - \bar z)} -
                 e^{-\frac i2 \sqrt{\mu} (z - \bar z)}\right]^{-2}\,,
\end{equation}
where $z=\phi+it/\ell$. Now remember that the asymptotic boundary of the
spacetime (\ref{SSdS}) has the topology of a cylinder. One can pass to the
plane (with coordinates $w, \bar w$) by the transformation $w=e^{iz}$,
leading to
\begin{equation}
e^{\gamma\Phi}\,dz d\bar z = \frac{\mu}{(w\bar w)^{1 - \sqrt{\mu}}\left[
                           1 - (w\bar w)^{\sqrt{\mu}}\right]^2}\, dw d\bar w\,,
\label{elliptic}
\end{equation}
which is the standard classical elliptic solution of Liouville
theory \cite{Seiberg:1990eb}.
Semiclassically, the Liouville vertex operators $e^{\alpha\Phi}$ appear
as sources of curvature in the classical equation of motion and lead
to solutions with local elliptic monodromy with $\sqrt{\mu} = 1 - \gamma
\alpha$ \cite{Seiberg:1990eb}. From this we obtain
\begin{equation}
\alpha = \frac{1 - \sqrt{\mu}}{\gamma}\,,
\label{relalphamu}
\end{equation}
i.~e.~, a relation between the mass parameter $\mu$ of the dS$_3$
solution and the parameter $\alpha$ of the vertex operator.
In the classical theory, $e^{\alpha\Phi}$ has conformal dimension
\begin{equation}
\Delta_{\mathrm{class}}(e^{\alpha\Phi}) = \frac{\alpha}{\gamma}
= \frac{1 - \sqrt{\mu}}{\gamma^2} = \frac{\ell}{8G}(1 - \sqrt{\mu})\,.
\label{confdimclass}
\end{equation}
Now the Schwarzschild-de~Sitter solution contains a pair of conical
defects at antipodal points on the spatial two-sphere, so only one of
these is inside the cosmological horizon. The bare mass of a conical
defect is \cite{Deser:1983dr,Klemm:2002ir}
\begin{equation}
m_{\mathrm{class}} = \frac{1-\sqrt{\mu}}{4G}\,. \label{baremass}
\end{equation}
Comparing this with (\ref{confdimclass}), one obtains
\begin{equation}
\ell m_{\mathrm{class}} = \Delta_{\mathrm{class}} +
                      \bar{\Delta}_{\mathrm{class}}\,,
\end{equation}
so the classical conformal weights reproduce exactly the mass of the
classical point particle in dS$_3$ that causes the conical defect.

The quantum dimension of the operator $e^{\alpha\Phi}$ is given
by \cite{Seiberg:1990eb}
\begin{equation}
\Delta(e^{\alpha\Phi}) = \frac{\alpha}{\gamma} - \frac 12 \alpha^2
                         = \frac{1 - \mu}{2\gamma^2}
                         = \frac{c}{24}(1 - \mu) = \frac{l}{16G}(1 - \mu)\,.
\end{equation}
Using $\bar{\Delta} = \Delta$ (the vertex operators are scalars),
and $\tilde c = c$, we obtain
\begin{equation}
\Delta + \bar{\Delta} = \ell M + \frac{c + \tilde c}{24}\,,
\end{equation}
where $M=-\mu/8G$ denotes the mass of the Schwarz\-schild-de~Sitter
solution (\ref{SSdS}) \cite{Spradlin:2001pw}.
This means that the sum of the {\it quantum} conformal dimensions
$(\Delta + \bar{\Delta})/\ell$ coincides (modulo a constant shift
of $(c+\tilde c)/24\ell$ coming from the transformation from
the cylinder to the plane) with the mass $M$ associated to the {\it classical}
geometry (\ref{SSdS}).
This mass includes the contribution of the gravitational field, and
should not be confused with the bare mass (\ref{baremass}) of the particles
that cause the conical defects.

Eq.~(\ref{relalphamu}) establishes thus a quantitative correspondence between
vertex operators in the boundary CFT and Schwarz\-schild-de~Sitter solutions
in the bulk. A further point that can be checked is the equality of the
Liouville stress tensor for elliptic solutions \cite{Seiberg:1990eb} and
the Brown-York energy-momentum tensor of (\ref{SSdS}) \cite{Klemm:2002ir}.

Let us now consider the gravity solutions with $\mu < 0$.
We will see that they also have a nice interpretation in
Liouville theory. Defining $E$ by $\sqrt{\mu} = iE\gamma$,
one gets for the classical Liouville solution corresponding to (\ref{SSdS})
with $\mu < 0$
\begin{equation}
e^{\gamma\Phi} dzd\bar z = \frac{E^2\gamma^2}{4w\bar w\sin^2\left(
                           \frac{E\gamma}{2}\ln w\bar w\right)}dwd\bar w\,,
\label{hyperbolic}
\end{equation}
which can also be obtained from (\ref{elliptic}) by analytical continuation.
As is well-known (cf.~e.~g.~\cite{Ginsparg:is} for a review), the hyperbolic
solution (\ref{hyperbolic}) corresponds in the semiclassical limit
$\gamma \to 0$ to the normalizable quantum states $\psi_E$ with momentum $E$,
i.~e.~, to the so-called macroscopic states. Formally, these states
can be associated to vertex operators $e^{\alpha\Phi}$ with
\begin{equation}
\alpha = \frac Q2 + iE\,. \label{alphaQE}
\end{equation}
If we use $\sqrt{\mu} = iE\gamma$ in our relation (\ref{relalphamu})
that connects vertex operators and bulk solutions, we get exactly
(\ref{alphaQE}). Furthermore, since the quantum state $\psi_E$ has
energy $E^2/2 + Q^2/8$ \cite{Seiberg:1990eb}, the sum of the energies
of $\psi_E$ (right-moving) and $\psi_{-E}$ (left-moving) is
\begin{equation}
E^2 + \frac{Q^2}{4} = \frac{1 - \mu}{\gamma^2} = \ell M +
\frac{c + \tilde c}{24}\,,
\end{equation}
which gives again the mass $M=-\mu/8G$.

Summarizing, one has thus the following picture: Gravity solutions
that have a temperature (i.~e.~, with $\mu \ge 0$) correspond to
vertex operators with $\alpha = (1 - \sqrt{\mu})/\gamma$, i.~e.~,
to non-normalizable or microscopic states.
Solutions with $\mu < 0$ correspond to normalizable or macroscopic
states with real momentum $E$, where $E$ is given by $\sqrt{\mu} = iE\gamma$.
Classical dS$_3$ gravity encodes therefore (at least some of) the
quantum properties 
of Liouville theory. However, in the reduction from the SL$(2,\bC)$
Chern-Simons 
theory (\ref{CSaction}) to the Liouville model (\ref{Liouvilleact}) only one
conformal boundary was taken into account. It would be interesting to see what
happens if one considers both boundaries. Some discussion on this can be
found in \cite{Balasubramanian:2002zh}.

In general it is an unsettled question if the CFT dual to de~Sitter gravity
resides on one boundary or on both. By considering two-point correlators
with one point on ${\cal I}^-$ and another on ${\cal I}^+$,
Strominger \cite{Strominger:2001pn} argues that one can identify the past
and future boundaries of de~Sitter
space by identifying points connected by null geodesics, so that the
holographic dual is a field theory on one $(D-1)$-dimensional boundary
rather than two. On the other hand, the authors of \cite{Balasubramanian:2002zh}
propose that the dual CFT should involve two disjoint, but possibly
entangled factors. In order to clarify these points, one would like
to have an explicit example of dS/CFT emerging directly from string theory.
If one wants to mimic the argumentation that lead to the AdS/CFT
correspondence \cite{Maldacena:1997re}, this requires the existence of
brane solutions that interpolate between flat space and de~Sitter space
(times some internal manifold). But, apart from the fact that such
brane solutions seem to be forbidden by the no-go theorems of
section \ref{dSstring}, dS vacua break all supersymmetries
(in conventional supergravity theories). This makes it questionable how
far one could trust a Maldacena-type argument in this case.

A further strong argument in favor of dS/CFT that we should mention
comes from the study of quasi-normal modes, the decaying
perturbations of de Sitter space. 
Remarkably enough, Abdalla et al. \cite{Abdalla:2002hg} showed that the
quasi-normal modes of scalar perturbations of de Sitter space are
contained in the spectrum of boundary CFT correlators. These
quasi-normal mode frequencies are given by (see also \cite{Du:2004jt}) 
\[
\omega=-\frac{i}{\ell}(2n+l+h_{\pm})\quad \mathrm{or}\quad 
\omega=-\frac{i}{\ell}(2n-l-D+3+h_{\pm})
\]
for natural $n$ and $l$, where $h_{\pm}=(D-1\pm\sqrt{(D-1)^2-4m^2\ell^2})/2$
are the conformal weights of the dual boundary operators.
When the conformal weights $h_{\pm}$ are
real, which is a necessary condition for the CFT to be unitary, the
scalar perturbations do not 
propagate in the bulk. By contrast, well defined quasi-normal modes
exist in de Sitter space for complex weights, when the CFT is
non unitary. The frequencies appear as poles in the Fourier transform
of the boundary correlator in static coordinates, see Eq.~\eqref{cylcorr}
below.    

Let us finally comment on some criticisms concerning the existence of
a dS/CFT correspondence that appeared in \cite{Dyson:2002nt}.
The authors of \cite{Dyson:2002nt} considered
a general finite closed system described by a thermal density
matrix, and a thermal correlator

\begin{equation}
F(t) = <{\cal O}(0) {\cal O}(t)>\,.
\end{equation}

It was then shown that the long time average of $F(t)F^{\ast}(t)$
is non-zero and positive, which leads to a contradiction with the dS/CFT
result in static coordinates \cite{Klemm:2001ea}\footnote{We only consider
the simple case of (2+1)-dimensional de~Sitter space and operators ${\cal O}$
that couple to bulk scalars of mass $m$.},

\begin{equation}
<{\cal O}(0,0) {\cal O}(t,\phi)> \sim \left[\cosh\frac t{\ell} - \cos\phi
\right]^{-h}\,, \label{cylcorr}
\end{equation}

where $h = 1 + \sqrt{1 - m^2\ell^2}$. (\ref{cylcorr}) is not the
standard thermal correlator, rather it is the two-point function
for dimension $(h,h)$ operators on a cylinder, whose length
(not circumference) is parametrized by the Euclidean time coordinate
$t$. (\ref{cylcorr}) behaves like

\begin{equation}
<{\cal O}(0,0) {\cal O}(t,\phi)> \sim e^{-ht/\ell} \label{asbehav}
\end{equation}

for large $t$.
Clearly the behaviour (\ref{asbehav}) would imply a zero
long-time average. Obviously the apparent contradiction
found in \cite{Dyson:2002nt}, which is based on the assumption that the
dual CFT is described by a thermal density matrix,
is resolved if the conformal field theory does
not encode the thermal nature of de~Sitter space. We would like to
point out here some arguments in favour of this.

First of all, the concept of assigning a temperature to de~Sitter
space is well-defined only in the static patches. However, the
past (and future-) boundary, where the CFT resides, lies outside the
static region.
In particular, the local Tolman temperature of de~Sitter space,

\begin{equation}
T(r) = \frac{1}{2\pi \ell\sqrt{1 - \frac{r^2}{\ell^2}}}
\end{equation}

formally becomes imaginary for $r>\ell$. It might thus be that
the conformal field theory on ${\cal I}^-$ does not capture
the thermal nature of de~Sitter space.
If this is true, and if the dual CFT nevertheless
accounts somehow for de~Sitter entropy, we do not expect this
entropy to be thermal\footnote{A non-thermal interpretation of de~Sitter
entropy in terms of a sort of Euclidean entanglement entropy was
proposed in \cite{Kabat:2002hj}.}.

We will now argue that in the Liouville approach discussed above,
dS entropy has indeed a non-thermal interpretation.
Let us start with the KPZ equation for gravitational dressing
of CFT operators with bare conformal weight $\Delta_0$ by vertex operators
$e^{\alpha\Phi}$ (cf.~e.~g.~\cite{Ginsparg:is}),

\begin{equation}
\alpha - \frac Q2 = -\sqrt{\frac 14 Q^2 - 2 + 2\Delta_0}\,.
\end{equation}

Setting $\Delta = 1 - \Delta_0$ yields

\begin{equation}
\alpha - \frac Q2 = -\sqrt{\frac 14 Q^2 - 2\Delta}\,,
\label{KPZDelta}
\end{equation}

which is of course the formula for the quantum conformal weights
$\Delta$ of vertex operators $e^{\alpha\Phi}$. Now observe that the
entropy of the Schwarz\-schild-dS$_3$ solution (\ref{SSdS})
is given by

\begin{equation}
S = \frac{\pi \ell\sqrt{\mu}}{2G} = \frac{4\pi \sqrt{\mu}}{\gamma^2}\,,
\end{equation}

and, using (\ref{relalphamu}) as well as the background charge $Q = 2/\gamma$, that

\begin{equation}
\alpha - \frac Q2 = -\frac{\sqrt{\mu}}{\gamma} = -\frac{S\gamma}{4\pi}\,.
\label{alphaQS}
\end{equation}

Inserting (\ref{alphaQS}) into (\ref{KPZDelta}), one obtains

\begin{equation}
S = 2\pi\sqrt{2Q^2\left(\frac{Q^2}{8} - \Delta\right)}\,.
\end{equation}

If we finally use the central charge $c = 3Q^2$ and $\bar{\Delta} = \Delta$,
we get

\begin{equation}
S = 2\pi\sqrt{\frac c6\left(\frac{c}{24} - \Delta\right)} +
    2\pi\sqrt{\frac c6\left(\frac{c}{24} - \bar\Delta\right)}\,.
\label{Cardy}
\end{equation}

(\ref{Cardy}) looks like the Cardy formula for the asymptotic level
density of conformal field theories, but actually it is not, because
the signs of the terms $\Delta$ and $c/24$ are interchanged. Rather,
we saw that (\ref{Cardy}) is the KPZ equation. Looking at (\ref{alphaQS}),
one also sees what de~Sitter entropy corresponds to in Liouville
theory. Using

\begin{equation}
iE = \alpha - \frac Q2
\end{equation}

for the Liouville momentum $E$ (cf.~(\ref{alphaQE})), one finally obtains

\begin{equation}
S = -\frac{4\pi i}{\gamma}E\,,
\label{SE}
\end{equation}

so that de~Sitter entropy is essentially Liouville momentum.
In particular, it has no statistical meaning in this approach.
This is reminiscent of Wald's Noether charge interpretation
of black hole entropy \cite{Wald:1993nt}.
Note that the Schwarz\-schild-dS$_3$ solution with $\mu > 0$
corresponds to imaginary momentum $E$, so that $S$ is real,
as it should be.

\section{Finite-dimensional Hilbert Space?}

\label{finitedim}

Banks and Fischler independently proposed
\cite{Banks:2000fe,Fischler:2000} (see also \cite{Banks:2001yp} and
\cite{Bousso:2000nf}) that the Hilbert space of quantum gravity on 
de~Sitter space is finite-dimensional, and argued that dS entropy is to be
interpreted as the logarithm of the total number of states in the Hilbert
space\footnote{Goheer, Kleban and Susskind \cite{Goheer:2002vf} assume only the
weaker condition that the spectrum is discrete, which follows from
finiteness of 
the entropy. Note that a finite entropy does not necessarily require a
finite-dimensional Hilbert space.}.
Let us briefly review some evidence for this proposal, which is
of course at odds with a dS/CFT correspondence.

The first argument comes essentially from classical general relativity.
In an unpublished work, Horowitz and Itzhaki showed that the classical
phase space of four-dimensional general relativity with asymptotically
de~Sitter boundary conditions both in the past and in the future, is
compact, if the energy-momentum tensor is that of homogeneous
matter \cite{HorItz}. It is well-known that a compact phase space
yields a finite-dimensional Hilbert space when quantized.
The second argument in favour of a finite number of states
goes as follows: Try to contradict the idea that asymptotically dS
spaces have a finite number of degrees of freedom. Consider e.~g.~a
spacetime which is dS$_D$ in the remote past. As the volume of space
(a $(D-1)$-sphere) is very large, one can easily impose initial conditions
that have a larger entropy than dS. However these initial
conditions will not lead to an asymptotically dS solution in the remote
future, rather the spacetime will finish in a big
crunch \cite{Banks:2000fe}. Let us explain this in more detail:
If we attribute to each source on the spacelike past boundary ${\cal I}^-$
a finite energy density, no matter how small, then for some finite number of sources,
the resulting spacetime cannot be asymptotically dS in past and future.
The energy density grows as we approach the point of maximal
contraction, and at some point black holes form with radius larger than
the putative radius of the dS sphere. In other words, we encounter a
cosmological singularity and do not asymptote to dS space\footnote{We are grateful to
Tom Banks for clarifying correspondence on this point.}.
In the AdS/CFT correspondence, the boundary values $\Phi_0$
of bulk fields $\Phi$ are sources for CFT operators $\cal O$, but, according
to what was said above, in dS we cannot put arbitrary sources. This seems thus
to be a problem for a possible dS/CFT correspondence.

Apparently the Banks-Fischler argument does not take into account
the gravitational backreaction of the sources which can be arbitrarily big.
The classical example is that of a point mass in general relativity, where
Arnowitt, Deser and Misner showed that for any bare mass gravity responds
in a way that makes its clothed mass vanish \cite{Arnowitt:1962hi}.
Thus, in order to complete the argumentation by Banks and Fischler,
one has to show that the negative gravitational interaction energy cannot
accomodate the excitation of an arbitrarily large number of matter degrees
of freedom.

Further evidence for a finite-dimensional Hilbert space
comes from the consideration of black holes in de~Sitter space.
Whereas in asymptotically flat or AdS spacetimes one can have black
holes with arbitrarily large entropies, this is not possible in de~Sitter spaces,
where the entropy of a black hole is bounded from above by the entropy
of the Nariai solution, which is the largest black hole that can fit
within the cosmological horizon\footnote{Something similar happens for black
holes in G\"odel-type universes \cite{Gimon:2003ms}, whose size is bounded from
above by the velocity of light surface, beyond which closed timelike curves appear.
This leads also to a maximal entropy.}. Thus one cannot have excitations with
arbitrarily large entropy. We shall discuss this point further in
section \ref{dSthermo}.

One might argue that a free matter quantum field theory on dS, renormalized
such that the vacuum expectation value of the stress tensor is zero,
has an infinite number of degrees of freedom, and the vacuum state
would support de~Sitter spacetime. However, the claim by Banks and Fischler is
that the {\it combined} system matter plus gravity is described by a finite-dimensional
Hilbert space. That free matter fields on dS give rise to a stable theory
seems to be doubtful in view of the perturbative results described in the
introduction. (Already a small perturbation $\lambda \phi^4$ of a free
scalar leads to an exponential production of particles and thus to
instabilities \cite{Myhrvold:1983hx}).

Note that the isometry group SO$(D,1)$ of dS$_D$ has no nontrivial finite-dimensional
unitary representations. If the Hilbert space $\cal H$ is finite-dimensional
then the dS isometry group cannot act on $\cal H$ \cite{Witten:2001kn}.

It is not clear how a finite-dimensional Hilbert space is compatible with
a dS/CFT correspondence. Yet there are some possible loopholes that we
will describe below.

First of all, the infinite dimension of the CFT Hilbert space might be cut
down by a constraint such as $L_0 + {\bar L}_0 = 0$\cite{Strompriv},
where $L_0$, ${\bar L}_0$ denote the Virasoro generators in the dS$_3$/CFT$_2$
correspondence. After all, the total energy in dS is zero, because spatial
sections are closed. (This is just the gravitational analogue of the usual
Gauss law in electrodynamics).

The second way out of the conundrum is due to Witten \cite{Witten:2001kn}. In AdS,
near the boundary $r \to \infty$, the metric behaves as
\begin{equation}
ds^2 \to dr^2 + e^{2r} d{\vec x}^2\,,
\end{equation}
where $d{\vec x}^2$ denotes the flat line element. This is generalized to
\begin{equation}
ds^2 \to dr^2 + e^{2r} g^{(0)}_{ij} dx^i dx^j\,,
\end{equation}
with $g^{(0)}_{ij}$ being an arbitrary metric on the boundary,
conformal to $d{\vec x}^2$. By considering dependence on $g^{(0)}_{ij}$,
one gets correlation functions of the stress tensor in the boundary CFT,
\begin{equation}
<T_{i_1 j_1}(x_1) \ldots T_{i_m j_m}(x_m)> = \frac{\delta}{\delta g^{(0)}_{i_1 j_1}(x_1)}\cdot
\ldots \cdot \frac{\delta}{\delta g^{(0)}_{i_m j_m}(x_m)}I(g^{(0)}_{ij})\,,
\end{equation}
where $I(g^{(0)}_{ij})$ is the bulk action for the field $g_{ij}$ with boundary
value $g^{(0)}_{ij}$. In de~Sitter space, we have two boundaries
$t \to \pm \infty$, where the metric approaches
\begin{equation}
ds^2 \to -dt^2 + e^{\pm 2t} d\Omega_{D-1}^2 \qquad {\mbox{for}} \quad
t \to \pm \infty\,.
\end{equation}
Here, $d\Omega_{D-1}^2$ denotes the round metric on S$^{D-1}$. To prepare an
initial or final state $\ket i$ or $\bra f$, pick a conformal metric
$g^{(i)}$ or $g^{(f)}$ on the sphere and require
\begin{equation}
ds^2 \to \left\{\begin{array}{l} -dt^2 + e^{-2t} g^{(i)}_{ij} dx^i dx^j\,, \quad
                                 t \to -\infty\,, \vphantom{\displaystyle\frac.1} \\
                                 -dt^2 + e^{+2t} g^{(f)}_{ij} dx^i dx^j\,, \quad
                                 t \to +\infty\,. \end{array} \right.
\end{equation}
Then the path integral for metrics with this asymptotics gives an ``observable"
$\left<f|i\right>$. (Witten calls this ``metaobservable", because the formulation
of $\left<f|i\right>$ requires a global view of ${\cal I}^-$ and ${\cal I}^+$,
and this is not available to any observer living in de~Sitter space).
$\left<f|i\right>$ is an $\infty \times \infty$ matrix, but it may have finite
rank. This finite rank is the dimension of the Hilbert space \cite{Witten:2001kn}.

A further loophole that we should mention is the possibility that only a
finite-dimensional subspace $\cal H$ of the CFT Hilbert space leads from a
dS geometry in the remote past to a dS geometry in the asymptotic future,
whereas states not contained in $\cal H$ might lead to a big crunch,
in accordance with the discussion above.

Last but not least, there is the proposal by G\"uijosa and Lowe \cite{Guijosa:2003ze},
who emphasized that dS/CFT should be formulated using unitary principal
series representations of the de~Sitter isometry group/conformal group,
as opposed to the standard highest-weight representations usually considered
in conformal field theory\footnote{This was also observed
in \cite{Balasubramanian:2002zh}.}. This avoids the problems associated with
the non-unitarity of the highest-weight representations that appear
in \cite{Strominger:2001pn}, but suffers from
the drawback that the principal series representations of SO$(D,1)$ are
infinite-dimensional, and so do not account for the finite gravitational
entropy of dS space in a natural way. For this reason the authors of
\cite{Guijosa:2003ze} proposed to replace the classical isometry group by a
$q$-deformed version, where $q$ is a root of unity. This was carried out for
dS$_2$ in \cite{Guijosa:2003ze} and for dS$_3$ in \cite{Lowe:2004nw}, and it
was found that the unitary principal series representations deform to
{\it finite-dimensional} unitary representations of the quantum group.

In order to understand a little bit more in detail what is going on,
let us ask the question how we would recognize a possible holographic
dual of gravity on de~Sitter spaces. The answer for AdS/CFT was initially
through the symmetries, which exactly matched. Thus we might try to recognize
the dS dual by looking at its symmetries, and requiring that they contain
the SO$(D,1)$ isometry group of dS$_D$. As this group coincides with the
Euclidean conformal group in one dimension lower, this would suggest that
the holographic dual is a Euclidean CFT in $D-1$ dimensions. The point is
now that most of this group does not preserve the region that is causally
accessible to an observer (the static region, shaded part of
figure \ref{penrose}) \cite{Goheer:2002vf}. This causal patch is preserved by
the subgroup $\bR \times$ SO$(D-1)$, where $\bR$ corresponds to time
translations 
(in the time $t$ of (\ref{metrstatic})), and SO$(D-1)$ is generated by
rotations that 
leave the horizon invariant. Therefore the holographic dual associated with an
{\it individual} observer should have the symmetry group $\bR \times$ SO$(D-1)$
rather than SO$(D,1)$, and is thus not a conformal field theory in
$D-1$ dimensions.
Banks constructed a toy model of such a ``dS quantum mechanics"
\cite{Banks:2003cg}, 
by using fuzzy spheres, which allow to realize the spherical geometry of the
horizon in a way that is compatible with a finite number of
states\footnote{Also 
Li \cite{Li:2001ky} has utilized fuzzy spheres for a hypothetical
description of dS quantum mechanics. Parikh and Verlinde \cite{Parikh:2004ux}
proposed as holographically dual theory a spin system which has finite dimension
(being a representation of SO$(D-1)$), but in which nevertheless there are
SO$(D,1)$-invariant probabilities.}.
His model was able to reproduce qualitatively the entropies of
the cosmological and the black hole horizon.

In conclusion, the idea is that
each {\it individual} observer has access only to a finite amount of degrees of
freedom associated with the corresponding holographic region (her
causal patch). 
There are {\it local}, observer-dependent holographic screens,
which coincide with the observer's horizon \cite{Bousso:1999cb}.
Interestingly enough, however, Bousso's recipe to construct holographic
screens yields a second possibility in the case of dS
space, namely a {\it global} screen located either at ${\cal I}^-$ or at
${\cal I}^+$ \cite{Bousso:1999cb}. Using such a global screen seems to
lead to dS/CFT. Moreover, the idea that individual observers can access
only a fraction of the total degrees of freedom is challenged by the
semiclassical validity of the Reeh-Schlieder property in de Sitter
space \cite{Bros:1998ik,Borchers:1998nw}. Thus while it is true that
the static vacuum has only the $\bR \times$ SO$(D-1)$ symmetry, the
geodesic observer can nevertheless explore the full Hilbert space by
means of local operations performed on a de Sitter invariant vacuum in
his/her causal patch.

\section{De~Sitter Thermodynamics}

\label{dSthermo}

\begin{figure}[ht]
\begin{center}
\includegraphics[width=0.4\textwidth]{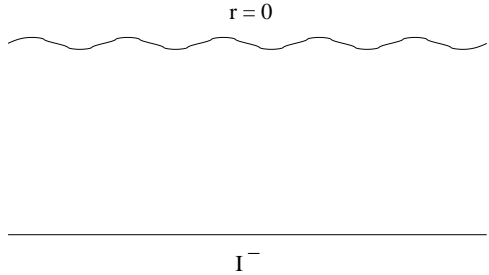}
\end{center}
\caption{\small{Carter-Penrose diagram for Schwarz\-schild-de~Sitter
space with mass parameter $m>m_N$. The solution starts from an
asymptotically de~Sitter geometry at past infinity ${\cal I}^-$
($r=\infty$), and finishes in a big crunch at $r=0$ (curvature
singularity). There is also a time inverted solution with a curvature
singularity in the past.}
}
\label{penr_naked}
\end{figure}

We conclude this review with a curiosity of de~Sitter
thermodynamics, namely the possible appearance of negative
absolute temperatures. The discussion below is not to
be intended as a proof, but rather as an alternative description
of the same physics.
Consider the Schwarz\-schild-dS$_4$
solution, which describes a black hole immersed in a dS background,
\begin{equation}
ds^2 = -V(r)dt^2 + V(r)^{-1}dr^2 + r^2 d\Omega_2^2\,, \label{SSdS4}
\end{equation}
with
\begin{equation}
V(r) = 1 - \frac{2m}r - \frac{r^2}{\ell^2}\,.
\end{equation}
For $0<m<m_N$, where
\begin{equation}
m_N = \frac{\ell}{3\sqrt 3}\,,
\end{equation}
there are two positive roots $r_h$ and $r_c>r_h$ where
$V(r)$ vanishes. $r_h$ corresponds to the black hole event horizon,
and $r_c$ to the cosmological horizon. For $m=m_N$ both roots
coalesce and we have the Nariai solution \cite{Nariai}, which
represents the largest black
hole one can have in de Sitter space. For $m<0$ the black hole
disappears, and the spacetime describes a naked singularity in $r=0$ surrounded
by a cosmological horizon. Finally, for $m>m_N$ there is no static region.
In this case the solution is asymptotically dS only in the far past
(cf.~the Carter-Penrose diagram, figure \ref{penr_naked}). If we want
to consider only spacetimes that approach dS in both past and future,
we have to discard the solutions with $m>m_N$.

The entropies associated to the black hole and the cosmological horizon
are given by $S_h=A_h/4G=\pi r_h^2/G$ and $S_c=A_c/4G=\pi r_c^2/G$
respectively, where $A_{h,c}$ denote the horizon areas.

It is well-known that the imaginary time periods required to avoid conical
singularities in the Euclidean section at both the black hole and the cosmological
horizons do not match \cite{Gibbons:1977mu}. Physically this corresponds to the fact
that the two horizons are not in thermal equilibrium.
Strictly speaking the only thermal equilibrium state is given by the Nariai
solution, with energy $E$ proportional to $m_N$.
Now start from this equilibrium state and subtract an infinitesimal amount of energy
from the system. This results in a small separation of the black hole and the
cosmological horizon. The black hole horizon shrinks slightly, whereas the
cosmological horizon increases. Hence the entropy of the black hole decreases with
decreasing energy, whereas the entropy of the cosmological horizon increases.
As we have
\begin{equation}
\frac 1T = \frac{\partial S}{\partial E}\,, \label{1/T}
\end{equation}
this behaviour implies that one should ascribe a negative temperature to
the cosmological horizon.

\begin{figure}[ht]
\begin{center}
\includegraphics[width=0.7\textwidth]{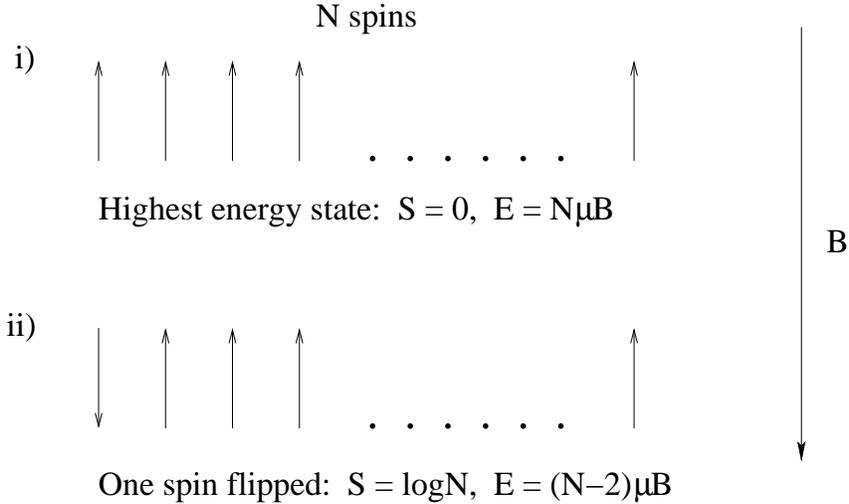}
\end{center}
\caption{\small{A typical system that exhibits negative temperatures:
$N$ atoms with spin $1/2$ and magnetic moment $\mu$ on a one-dimensional
wire, in an external magnetic field $B$ pointing down, with spin-flip
the only degree of freedom.}
}
\label{spins}
\end{figure}

At this point, let us open a parenthesis on negative absolute temperatures.
Negative absolute temperatures \cite{Ramsey:1956} occur whenever the entropy of a
thermodynamical system is not a monotonically increasing function of its internal energy.
As $T$ is defined by Eq.~(\ref{1/T}), we see that $T<0$ if
the entropy $S$ decreases with increasing $E$.
As a concrete example consider e.~g.~$N$ atoms with spin $1/2$ on a
one-dimensional wire in an external magnetic field pointing down (cf.~figure
\ref{spins}). Suppose that spin-flip is the only degree of freedom. In the highest energy
state all spins point up, and the entropy is zero (fig.~\ref{spins}i)).
If we flip one spin, the energy is lowered, but the entropy increases, because there
are $N$ available microstates (one of which is shown in fig.~\ref{spins}ii)).

The temperature is thus negative in this regime. Nuclear spin systems in pure LiF
crystals do indeed realize negative temperatures experimentally \cite{Purcell:1951}.
In these systems, the spin-lattice relaxation times are as large as $5$ minutes,
whereas the spin-spin relaxation times are less than $10^{-5}$ seconds, so that
the defintion of a spin temperature makes sense. Another well-known example displaying
negative temperature is the laser (population inversion). A further class of problems
that exhibit negative temperatures is the statistical mechanics of a vortex gas or
Coulomb gas in two dimensions. Notice that any negative temperature is hotter than
any positive temperature while for two temperatures of the same sign the one with
the algebraically greater value is the hotter \cite{Ramsey:1956}.

Note also that a necessary condition for the appearance of negative temperatures is
an upper bound to the possible energy of the allowed states. To see this,
consider the Boltzmann factor $\exp(-E_n/k_B T)$, which increases exponentially
with increasing $E_n$ for $T < 0$, so that the high-energy states are occupied
more than the low-energy ones. (As already mentioned above, this happens
e.~g.~in a laser). Consequently, with no upper limit to the energy,
negative temperatures could not be achieved with a finite energy.

But this is exactly what happens in de~Sitter gravity, where the mass is bounded
from above by the mass of the largest black hole that can fit within the
cosmological horizon. Furthermore, negative temperatures typically occur in
systems with finite-dimensional Hilbert spaces (like the spin system that we
discussed). The observed thermodynamical behaviour fits thus with the claim
by Banks and Fischler.

Actually, for the Schwarz\-schild-de~Sitter black hole (\ref{SSdS4}),
even the {\it total} entropy of the black hole and the cosmological horizon decreases
with increasing energy. This can be shown by using the result of
Gibbons and Hawking \cite{Gibbons:1977mu}, who integrated the Killing identity on a
spacelike hypersurface $\Sigma$ from $r_h$ to $r_c$ to get the Smarr-type
formula
\begin{equation}
GM_c = \frac{\kappa_h A_h}{4\pi} + \frac{1}{4\pi} \int_{\Sigma} \Lambda K_{\mu}\, d\Sigma^{\mu}\,,
      \label{Smarr}
\end{equation}
where $M_c = -\kappa_c A_c/4\pi$ denotes the total mass within the cosmological
horizon and $\kappa_{h,c}$ are the surface gravities of the black hole and the
cosmological horizon respectively. Furthermore, $K = \partial_t$ and $\Lambda = 3/\ell^2$
is the cosmological constant. One can interpret the first term on the rhs of
Eq.~(\ref{Smarr}) as the (positive) mass of the black hole, and the second term as
the (negative) contribution of $\Lambda$ to the total mass $M_c$ within the cosmological
horizon \cite{Gibbons:1977mu}.
Evaluating (\ref{Smarr}) yields
\begin{equation}
GM_c = \frac{r_c}{2} - \frac{3r_c^3}{2\ell^2}\,.
\end{equation}
Using the relation
\begin{equation}
r_h^2 + r_c^2 + r_h r_c = \ell^2\,,
\end{equation}
as well as $r_c \ge \ell/\sqrt 3$ (the minimum value of $r_c$ is obtained for the Nariai
black hole), it is straightforward to show that $M_c$ is a monotonically decreasing
function of the total entropy $S_{tot} = S_h + S_c = \pi(r_h^2 + r_c^2)/G$.
If we start from pure de~Sitter space and form a black hole, then the total mass
within the cosmological horizon increases, but this excitation has lower total entropy
than dS space itself.

In three dimensions, the black hole horizon degenerates to a conical singularity
in $r=0$, and the Schwarz\-schild-de~Sitter solution with mass $E$
reads \cite{Deser:1983dr}
\begin{equation}
ds^2 = -\left(1 - 8GE - \frac{r^2}{\ell^2}\right)dt^2 + \left(1 - 8GE - \frac{r^2}{\ell^2}
       \right)^{-1}dr^2 + r^2 d\phi^2\,.
\end{equation}
The entropy of the cosmological horizon is $S_c = \pi r_c/2G$, where
$r_c^2 = \ell^2(1 - 8GE)$. This leads to the
thermodynamic fundamental relation
\begin{equation}
E(S_c) = \frac 1{8G} [1 - (2GS_c/\pi \ell)^2]\,,
       \label{ED=3}
\end{equation}
and thus $\partial E/\partial S_c$ yields minus the temperature normally assigned
to dS space. In order to explain this, the authors of \cite{Spradlin:2001pw} argued
that instead of (\ref{1/T}) one should use
\begin{equation}
\frac 1T = \frac{\partial S}{\partial(-E)} \label{modified}
\end{equation}
to compute the dS temperature. The reason for this is that de~Sitter entropy is
supposed to correspond to the entropy of the degrees of freedom behind the horizon which cannot
be observed. As the spatial sections of dS are spheres, putting something with
positive energy on the north pole (where we assume that our observer sits)
implies that necessarily there will be some
negative energy on the south pole (i.~e.~, beyond the horizon of our observer),
therefore the minus sign in Eq.~(\ref{modified}).
If the observer instead varies the entropy with respect to the energy $+E$ within her
horizon then the usual laws of thermodynamics apply, but the price to pay is the
introduction of a negative temperature.

\section*{Acknowledgements}
\small

This work was partially supported by INFN, MURST and
by the European Commission RTN program
HPRN-CT-2000-00131, in which D.~K.~is
associated to the University of Torino. We would like to thank
V.~Balasubramanian, T.~Banks and A.~Strominger for useful
discussions/correspondence.

\end{document}